\documentclass[aps,prb,twocolumn,superscriptaddress,showpacs]{revtex4}

\usepackage{graphicx}
\usepackage{dcolumn}
\usepackage{bm}

\newcommand{\be}{\begin{equation}}
\newcommand{\ee}{\end{equation}}
\newcommand{\bd}{\begin{displaymath}}
\newcommand{\ed}{\end{displaymath}}
\newcommand{\ba}{\begin{eqnarray}}
\newcommand{\ea}{\end{eqnarray}}
\newcommand{\nn}{\nonumber \\}
\newcommand{\bbar}{b^+}
\newcommand{\cbar}{c^+}
\newcommand{\fbar}{f^+}

\newcommand{\ek}{\epsilon_k}
\newcommand{\ebk}{\epsilon^b_k}
\newcommand{\efk}{\epsilon^f_k}
\newcommand{\Efk}{E^f_k}
\newcommand{\Ebk}{E^b_k}
\newcommand{\Ebq}{E^b_q}
\newcommand{\kbar}{\overline{k}}
\newcommand{\ij}{\langle ij \rangle}

\newcommand{\chib}{\chi^b}
\newcommand{\chif}{\chi^f}
\newcommand{\Deltab}{\Delta^b}
\newcommand{\Deltaf}{\Delta^f}

\begin{document}

\title{Phases of Mott-Hubbard bilayers}

\author{Jung Hoon Han}
\email[Electronic address:$~$]{hanjh@skku.edu}
\affiliation{Department of Physics and Institute for Basic Science Research, \\
Sungkyunkwan University, Suwon 440-746, Korea}
\affiliation{CSCMR, Seoul National University, Seoul 151-747, Korea}
\author{Chenglong Jia}
\email[Electronic address:$~$]{cljia@skku.edu}
\affiliation{Department of Physics and Institute for Basic Science Research, \\
Sungkyunkwan University, Suwon 440-746, Korea}
\date{\today}

\begin{abstract}
A phase diagram of two Mott-Hubbard planes interacting with a
short-range Coulomb repulsion is presented. Considering the case of
equal amount of doping by holes in one layer as electrons in the
other, a holon-doublon inter-layer exciton formation is shown to be
a natural consequence of Coulomb attraction. Quasiparticle spectrum
is gapped and incoherent below a critical doping $\delta_c$ due to
the formation of excitons. A spin liquid insulator (SLI) phase is
thus realized without the lattice frustration. The critical value
$\delta_{c}$ sensitively depends on the inter-layer interaction
strength. In the $tJ$ model description of each layer with the
$d$-wave pairing, $\delta_{c}$ marks the crossover between SLI and
$d$-wave superconductor. The SLI phase, despite being
non-superconducting and charge-gapped, still shows electromagnetic
response similar to that of a superfluid due to the exciton
transport. Including antiferromagnetic order in the $tJ$ model
introduces magnetically ordered phases at low doping and pushes the
spin liquid phase to a larger inter-layer interaction strength and
higher doping concentrations.
\end{abstract}

\maketitle

\section{Introduction}

A realization of atomically sharp interface of a band insulator
(SrTiO$_3$) and a Mott insulator (LaTiO$_3$) and the observation of
a metallic interfacial layer between them\cite{hwang} has prompted a
lot of theoretical work on the interfacial phenomena when two
insulating materials of completely different physical origin are
joined together\cite{millis,others}.  In light of rapid experimental
advances in the growth of digitally sharp layered materials, it
seems likely that the successful construction of atomically sharp
interface between two (doped) Mott insulators is not very far. In
fact, a successful construction of alternating LCO:LSCO atomic
layers was demonstrated recently\cite{bozovic}. A well-known fact
that some high-$T_c$ materials contain bilayers of doped Hubbard
planes adds to the practical importance of studying the bilayer Mott
system theoretically. In these regards, a simple yet reliable
calculation scheme for understanding the behavior of heterostructure
consisting of Mott-insulating materials is called for.  We provide
one such theoretical framework in this paper. A closely related work
considering a correlated bilayer model can be found in Ref.
\onlinecite{berkeley}.

The situation we have in mind is that of two lightly-doped
conducting planes subject to Hubbard-$U$ repulsion, coupled to each
other through short-range Coulomb interaction. Each plane is modeled
as the Hubbard model, or its large-$U$ equivalent as the $tJ$ model.
In particular we focus on the case of symmetric \textit{p-n} doping,
with  density $\delta$ of holes in one layer and an equal amount of
doubly occupied electrons (doublons) in the other. The case
$\delta=0$ corresponds to decoupled, half-filled Mott insulators in
each layer.

The effective holon-doublon attraction results in the exciton
binding, which pervades the phase diagram and results in a number of
unusual properties that have no analogs in the single-layer model.
For weak inter-layer Coulomb interaction, phases found for a
single-layer model have their counterparts in the bilayer system.
Due to the exciton binding, however, an incoherent regime
characterized by a gap in the charge excitation and ill-defined
quasiparticle peaks dominates the low-doping region. This phase is
christened the ``spin-liquid-insulator" (SLI) phase. The in-plane
electromagnetic response in the SLI regime are shown to be
influenced by the exciton gap; in particular a superfluid response
is found within each layer even though the overall phase is not that
of a superconductor. When each plane is modelled as the $tJ$ model
with long-range magnetic order, the low-doping region is shown to be
dominated by antiferromagnetism and the SLI phase is found for a
large inter-layer Coulomb interaction and doping concentration,
leading to an interesting possibility to realize a spin liquid
insulator state without the lattice frustration.

Since our attempt is quite new, we present the analysis of the
bilayer model in increasing degree of complexity starting with the
$U\rightarrow \infty $ limit of the Hubbard model in section
\ref{Hubbard-plane}. Basic strategy and formalism for the analysis
of the bilayer physics is laid out. Hints of new phenomena unique to
the bilayer, as opposed to the same model in a single layer, are
pointed out. In section \ref{non-magnetic-t-J}, we consider each
plane described by the $tJ$ model with the possibility of having a
$d$-wave pairing of the electrons. Finally, antiferromagnetic
ordering which was ignored in section \ref{non-magnetic-t-J} is
included and its results are compared with the non-magnetic $tJ$
model in section \ref{magnetic-t-J}. The inter-plane coupling is
assumed to be the short-range Coulomb interaction throughout the
paper. Single-electron hopping across the layers is not considered.
We discuss the overall phase diagram and some spectral as well as
electrical properties of the phases found with particular emphasis
on those aspects which have no counterparts in the single layer.

\section{Hubbard Model Formulation}
\label{Hubbard-plane}

In this section we outline the basic formulation and solve the
bilayer model when each plane is characterized by the Hubbard model
with $U$ taken to infinity. This corresponds to the limit where only
the constrained hopping of electrons is available. Each plane is
assumed identical, having the same Hubbard interaction $U$ and the
kinetic energy scale $t$. The interaction between the planes is
mediated by a Coulomb repulsion ($n_{ai} = \cbar_{ai}c_{ai}$)

\bd H_{12} = V \sum_i (n_{1i}-1)(n_{2i}-1) \ed
for each lattice site $i$. The two planes are designated by $a =
1,2$. The Hubbard model for plane $a$ is

\be H_a = -t \sum_{ij \sigma} \cbar_{aj\sigma}c_{ai\sigma} + {U\over
2} \sum_i (n_{ai}-1)^2 -\mu_a \sum_i n_{ai}
.\label{Hubbard-model}\ee 
The Hamiltonian for the combined system reads
\be H = H_1 + H_2 + H_{12} . \ee
We consider the case of a symmetric chemical potential mismatch
$\mu_1 =\mu = -\mu_2 $ and $\mu>0$, which leads to hole doping for
$a=1$ and the doping by an equal amount of doublons in the second
layer.

\begin{figure}[t]
\begin{center}
\includegraphics[angle=0,width=6cm]{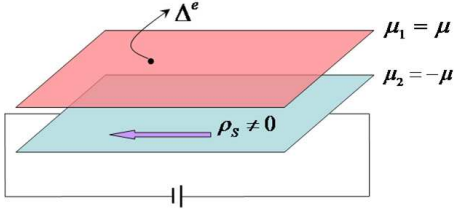}
\end{center}
\caption{{\protect\small (color online) Schematic plot of the two
Mott-Hubbard bilayer configuration. The chemical potential is chosen
to have the same amount of holons in one layer as the doublons in
the other. In-plane excitations in this model for doping $\delta <
\delta_c$ (see text for definition of $\delta_c$) is depicted.
Single-particle excitation requires a finite energy $\Delta^e$ due
to the exciton gap. In-plane electromagnetic response is that of a
superfluid even when the whole system is not a superconductor.}}
\label{response-of-bilayer}
\end{figure}
%

It proves convenient to first carry out the following particle-hole
transformation of the doublon ($a=2$) layer,
\ba c_{2i\sigma} &\rightarrow& (-1)^i \cbar_{2i\sigma}  \nn
    n_{2i} &\rightarrow&  2-n_{2i}, \label{ph-transform}\ea
to re-write
\ba H_2 &\rightarrow & -t\sum_{ij\sigma}
\cbar_{2j\sigma}c_{2i\sigma} + {U\over2}\sum_i (n_{2i}-1)^2 -\mu
\sum_i n_{2i}  \nn
   H_{12} & \rightarrow & -V \sum_i (n_{1i}-1)(n_{2i}-1) .\ea
Due to the particle-hole transformation for $a=2$, both layers now
appear as hole-doped and the inter-layer Coulomb interaction becomes
attractive.

Double occupation of a given site is prohibited in the large-$U$
limit, and it is convenient to introduce the slave-boson
representation of the electron operator
\ba c_{ai\sigma} &=& \bbar_{ai} f_{ai\sigma},  \nn n_{ai} &=& 1-
\bbar_{ai}b_{ai} . \ea
The constraint $\bbar_{ai} b_{ai} + \sum_\sigma \fbar_{ai\sigma}
f_{ai\sigma} = 1 $ is included via Lagrange multiplier
$\lambda_{ai}$. The slave-boson representation results in
constrained hopping model for each layer:

\ba H_a &=& -t \sum_{ij\sigma} \bbar_{ai}b_{aj}
\fbar_{aj\sigma}f_{ai\sigma} +\mu \sum_i \bbar_{ai}b_{ai} \nn &&
+\sum_i \lambda_{ai} (\bbar_{ai}b_{ai} +\sum_\sigma
\fbar_{ai\sigma}f_{ai\sigma} -1 ) \ea
while the inter-layer interaction takes on

\be H_{12} = -V \sum_i \bbar_{1i}b_{1i}\bbar_{2i}b_{2i} .
\label{transformed-Hubbard}\ee

The resulting Hamiltonian $H= H_1 + H_2 + H_{12}$ is solved in the
slave-boson mean-field theory  using the following self-consistent
parameters
\ba && \chif = \sum_\sigma \langle \fbar_{1j\sigma} f_{1i\sigma}
\rangle = \sum_\sigma \langle \fbar_{2j\sigma} f_{2i\sigma} \rangle
\nn
&& \chib = \langle \bbar_{1j} b_{1i} \rangle= \langle \bbar_{2j}
b_{2i} \rangle  \nn
&& \Deltab = \langle b_{1i}b_{2i} \rangle .\label{OP}\ea
All order parameters are assumed real and uniform, and $\lambda_{ai}
=\lambda $ is also assumed constant, uniform, and equal for both
planes. Non-zero $\Deltab$ signifies the presence of excitons formed
of doublon-holon pairs across the layers. The mean-field Hamiltonian
thus obtained is a sum of bosonic and fermionic parts which reads,
in momentum space,
\ba H^f_{mf} &=& \sum_{ak\sigma } \efk \fbar_{ak\sigma}f_{ak\sigma}
\nn
H^b_{mf} &=& \sum_{ak} \ebk  \bbar_{ak}b_{ak} - V\Deltab \sum_k
(\bbar_{1\kbar}\bbar_{2k} + b_{1\kbar}b_{2k} ), \nn
\label{mf-H}\ea
where $\kbar= -k$. We introduced the shorthand notation $\efk =
\chib \ek + \lambda$ and $\ebk = \chif \ek + \mu+\lambda$ for the
renormalized fermion and boson dispersions, where $\ek = -2t [\cos
k_x + \cos k_y ]$ is the bare band dispersion appropriate for a
square lattice. For later convenience we re-define $\mu$ in such a
way that the boson dispersion becomes

\bd \ebk = \chif (\ek + D) +\mu \ed
with $D = 4t$ being the half bandwidth of the bare band. After
carrying out appropriate Bogoliubov rotation the boson Hamiltonian
is diagonalized with the quasiparticle energy $\Ebk = \sqrt{(\ebk
)^2 -(V\Deltab )^2 }$.

There are two equations one can write down for the fermion and boson
occupation numbers respectively, and another three for the three
order parameters introduced earlier in Eq. (\ref{OP}). In these
equations we replace the momentum sum $\sum_k$ by an integral $\int
d\epsilon D(\epsilon)$ and use the simplistic form of the density of
states $D(\epsilon) = 1/(2D)$. This approximation allows us to solve
the self-consistent equations analytically. Without going through
the rather straightforward derivations in detail, we quote the final
results of solving the five equations at zero temperature:

\begin{widetext}
\ba && \lambda = \delta \times \chib D \nn
&& {1\over2} +\delta = {1\over 4\chif D} \left( \sqrt{(2\chif D +
\mu)^2 - (V\Deltab)^2 } -\sqrt{\mu^2 - (V\Deltab)^2} \right) \nn
&& \chif = (1-\delta^2 )/2 \nn
&& \chib = {1\over 2\chif D} \Bigl( (\mu + \chif D )(\delta +
{1\over 2}) - V(\Deltab)^2 - {1\over 4} \bigl(\sqrt{(2\chif D +
\mu)^2 - (V\Deltab)^2 } +\sqrt{\mu^2 - (V\Deltab)^2}   \bigr) \Bigr)
\nn
&& \Deltab = {1\over 2} V\Deltab \ln \left( { {1 + \mu + \sqrt{(1 +
\mu)^2 - (V\Deltab)^2 } } \over { { \mu + \sqrt{\mu^2 - (V\Deltab)^2
}}}} \right). \label{SC-eq-I}\ea
\end{widetext}
The solution of Eq. (\ref{SC-eq-I}) gives two phases, at $T=0$, as
the doping $\delta$ is varied. For $\delta < \delta_c$ a featureless
state is obtained with a gap in the single-particle spectrum. There
is no well-defined quasiparticle peak in this regime. This phase is
dubbed the spin liquid insulator, or SLI. For $\delta
> \delta_c$ the gap closes, quasiparticle coherence
recovered, and one is in the Fermi liquid (FL) regime. The critical
doping $\delta_c$ which marks the incoherent-to-coherent crossover
is zero for $V=0$ and grows as $V$ is increased. In the following we
discuss the derivation of the results mentioned above. For
convenience we set the overall energy scale $2\chif D$ to unity.

From the second of Eq. (\ref{SC-eq-I}) one obtains

\be (1+2\delta)^2 [\mu^2- (V\Deltab)^2 ] = [\mu- 2\delta
(1+\delta)]^2 .\label{21}\ee
It is clear that self-consistency is only fulfilled for $\mu \geq
V\Deltab$, whereas $\mu = V \Deltab$ marks the onset of Bose
condensation, or BEC. Exact expressions for $\mu$ and $V\Deltab$ can
be found as

\ba \mu &=& {1+\delta + \delta e^{2/V} \over e^{2/V}-1} \nn
 V\Deltab &=& {2\sqrt{\delta(1+\delta)}e^{1/V} \over e^{2/V}-1}.
\label{23}\ea
Note that $\mu \geq V\Deltab $ according to (\ref{23}). The two
values coincide at the critical doping given by (restoring $2\chif
D$ temporarily)

\be \delta_{c} = {1\over e^{4\chif D /V} -1} .\label{delta-BEC}\ee
The exciton amplitude $\Deltab$ grows as the square root of the
hole(doublon) density for small doping. The fermion part of the
mean-field Hamiltonian carries no gap, but the boson spectrum has a
gap equal to
\be \mathrm{min}(\Ebk ) = \sqrt{\mu^2 - (V\Deltab)^2} = { 1+\delta
-\delta e^{2/V} \over e^{2/V}-1 }.\label{min-Ek}\ee
The gap vanishes at $\delta = \delta_{c}$.

When $\delta$ exceeds $\delta_{c}$, Bose condensation guarantees
$\mu$ is fixed to $V\Deltab$. For $\delta > \delta_{c}$, the
residual boson density $\delta-\delta_{c}$, defined as the
condensate fraction, goes into the condensate portion of the boson
state. The spectral weight of the coherent quasiparticle in the
$\delta > \delta_c$ regime grows as $\delta-\delta_c$. The critical
doping depends on the interaction strength $V$ in the manner given
in Eq. (\ref{delta-BEC}). It is zero for $V=0$ as expected for a
single-layer Hubbard model.

The $T=0$ Green's function for the electron in the SLI phase,
$\delta < \delta_{c}$, can be worked out

\ba G^e (k,i\omega) &=& \sum_q \cosh^2 \theta_q { \theta (
-\epsilon^f_{k+q} ) \over \epsilon^f_{k+q} - E^b_q - i\omega} \nn
&+& \sum_q \sinh^2 \theta_q { \theta(\epsilon^f_{k+q} ) \over
\epsilon^f_{k+q} + E^b_q - i\omega } \nn \ea
where
\bd \cosh^2 \theta_q = {1\over2} \left( {\epsilon^b_q \over \Ebq }
+1 \right) , ~ \sinh^2 \theta_q = {1\over2} \left( {\epsilon^b_q
\over \Ebq } -1 \right)  .\ed
Momentum-integrated spectral function $A^e (\omega)$ is obtained as
the imaginary part of the quantity $\sum_k G^e (k, i\omega
\rightarrow \omega + i\delta)$. Evaluating the sum over $k$ as an
integral with a constant density of states as before, we get the
spectral function $A^e (\omega)$:

\ba && {1\over 2D} \sum_q \cosh^2 \theta_q ~\theta (-\Ebq - \omega)
\int_{-D}^D \delta ( \chib \epsilon - \Ebq - \omega) d\epsilon \nn
&& + {1\over 2D} \sum_q \sinh^2 \theta_q ~\theta (\omega -\Ebq )
\int_{-D}^D \delta ( \chib \epsilon + \Ebq - \omega) d\epsilon .\nn
\ea
The step function $\theta(-\omega - \Ebq )$ appearing in the first
line is non-zero only if $\omega < -\Ebq$. The step function in the
second line is non-zero only if $\omega > \Ebq$. Therefore, in the
range where $|\omega | < \mathrm{min}(\Ebq)$ we get $A^e (\omega) =
0$. The gap in the electronic spectral function matches that in the
boson spectrum obtained in Eq. (\ref{min-Ek}), establishing $\delta
<\delta_c$ as the charge-gapped phase.

Despite the presence of the gap in the single-particle sector the
SLI phase is not an electronic insulator, but a metal. It can be
demonstrated by referring to the Ioffe-Larkin
rule\cite{ioffe-larkin}, which states that the conductivity
$\sigma^e$ of the fermion-boson composite system is given as $
(\sigma^e )^{-1} = ( \sigma^b )^{-1} + (\sigma^f )^{-1}$.
As will be shown in the next section, the exciton pairing has a
similar influence on the electromagnetic response of the bosons as
the BCS pairing on fermions, and the boson conductivity $\sigma^b$
in the SLI phase is infinite.  The fermion part is metallic,
however, and the overall response of the electron system is that of
a metal according to Ioffe-Larkin rule. The peculiarity arises from
the bilayer nature of the model. Extracting a charge from an
individual layer costs an energy gap because one has to break the
exciton pair. On the other hand, in-plane motion of a holon is
always accompanied by that of a doublon, which together make up an
exciton that moves without an energy cost.

For $\delta > \delta_{c}$ condensation of single bosons allows us to
write the coherent part of the spectral function as $A^e (\omega )
\sim (\delta - \delta_{c} ) A^f (\omega)$ and $A^f (\omega)$ is
constant near $\omega = 0$ as in the non-interacting Fermi liquid.
Therefore we have a metallic state with coherent quasiparticles for
$\delta > \delta_{c}$. One can say that $\delta_{c}$ marks an
incoherent-coherent crossover of the quasiparticles, while the
electric transport property remains  metallic for all non-zero
doping.

Analysis for the second Hubbard layer is similar. Due to the
particle-hole transformation (\ref{ph-transform}) the electron
Green's function $\langle T_\tau c_{2i} (\tau) \cbar_{2j} \rangle$
in the original model is mapped onto $-(-1)^{i-j} \langle T_\tau
\cbar_{2j} (\tau) c_{2j} \rangle = -(-1)^{i-j} G^e_{2ji} (-\tau)$
after the transformation. Likewise, the electronic Green's function
at $(k,\omega)$ is obtained as $G^e (Q-k, -\omega)$, $Q=(\pi,\pi)$,
in the transformed Hamiltonian. However, the argument in regard to
the energy gap is done summing over all available momenta $k$, so
the distinction between $k$ and $Q-k$ vanishes. The second layer has
a gap with the same magnitude as the first layer for $\delta <
\delta_{c}$, has metallic conduction, etc.
\\

\section{ t-J Model Formulation}
\label{non-magnetic-t-J}

Instead of the Hubbard model in Eq. (\ref{Hubbard-model}) we adopt
the $tJ$ model for each plane:

\be H_a = -t \sum_{ij \sigma} \cbar_{aj\sigma}c_{ai\sigma} -\mu_a
\sum_i n_{ai} +J \sum_{\ij} S_{ai}\cdot S_{aj}.\ee
A well-known mapping of the Hubbard model to the $tJ$ model for
large $U/t$ justifies the use of $tJ$ model for the doped Mott
planes. The spin-exchange term introduces additional degrees of
freedom having to do with $d$-wave pairing of electrons and
long-range magnetic order. We will deal with the first possibility
in this section and consider the magnetic order in the next.

Under the particle-hole transformation (\ref{ph-transform}) for the
second layer the spin operator $(S^x , S^y, S^z )$ is transformed to
$(-S^x, S^y, -S^z )$, but the inner product $S_i \cdot S_j $ remains
unaffected. The Hamiltonian, after the particle-hole transformation
is carried out, is re-written using the slave-boson substitution as
the mean-field type $H_{mf} = H^b_{mf} + H^f_{mf}$ by introducing
the mean-field parameters $\chib_{ij} = \langle \bbar_{aj}b_{ai}
\rangle$, $\chif_{ij} = \sum_\sigma \langle
\fbar_{aj\sigma}f_{ai\sigma} \rangle$, $\Deltab_i = \langle
b_{1i}b_{2i} \rangle$, and $\Deltaf_{ij} = \langle
f_{ai\uparrow}f_{aj\downarrow}-f_{ai\downarrow}f_{aj\uparrow}\rangle$:

\ba  H^f_{mf}  &= &- \sum_{aij \sigma} (t\chib_{ji} + J\chif_{ji})
\fbar_{aj\sigma}f_{ai\sigma} +\sum_i \lambda_{ai}
\fbar_{ai\sigma}f_{ai\sigma}\nn
&&  ~~~- {J} \sum_{a\ij} ( \Deltaf_{ij} P^+_{aij}  + h.c.) \nn
 H^b_{mf} &=& -\sum_{aij}t \chif_{ji}\bbar_{aj} b_{ai} +
\sum_{ai}(\mu + \lambda_{ai} ) \bbar_{ai}b_{ai} \nn && - V\sum_i
\Deltab_i (\bbar_{1i}\bbar_{2i}+b_{1i}b_{2i}). \label{tJ-model}\ea
The fermion pairing operator is given by
$P_{aij}=f_{ai\uparrow}f_{aj\downarrow}-f_{ai\downarrow}f_{aj\uparrow}$
in $H^f_{mf}$. Assuming all mean-field parameters are real and
uniform and that the fermion gap obeys the $d$-wave symmetry, we can
write down

\ba H^f_{mf} &=& \sum_{ak\sigma } \efk \fbar_{ak\sigma}f_{ak\sigma}
- \sum_{ak} \Deltaf_k [f_{ak\uparrow}f_{a\kbar \downarrow} +
\fbar_{a\kbar \downarrow}\fbar_{ak\uparrow} ] \nn
H^b_{mf} &=& \sum_{ak} \ebk  \bbar_{ak}b_{ak}  - V\Deltab \sum_k
[\bbar_{1\kbar}\bbar_{2k} + b_{1\kbar}b_{2k} ] \nn
\efk &=& \left( \chib  + {J\chif \over t } \right) \ek + \lambda \nn
\ebk &=& \chi_f (\ek + D ) + \mu \nn \ek &=& -2t (\cos k_x + \cos
k_y ) \nn
\Deltaf_k &=& 2J \Deltaf (\cos k_x - \cos k_y ) .\label{mf-H2}\ea
The self-consistent parameters $\Deltaf$ and $\chif$ follow from

\ba &&\Deltaf =  J \Deltaf\sum_k {(\cos k_x - \cos k_y)^2 \over \Efk
}  \nn
&& \chif =  -{1\over2}\sum_k (\cos k_x + \cos k_y ) {\efk \over \Efk
}  \nn \label{chif-eq} \ea
at zero temperature. The fermion spectrum is diagonalized with $\Efk
= \sqrt{(\efk )^2 + (\Deltaf_k )^2 }$. At zero doping $\efk$ is
reduced to $-2J\chif (\cos k_x + \cos k_y)$ and one readily obtains
$\chif =\Deltaf\approx 0.33J $, as expected from SU(2) symmetry and
in agreement with Kotliar and Liu's earlier
calculation\cite{kotliarliu}. The hole number in each layer is given
by $\delta = \sum_k  (\efk / \Efk )$.

Notice that all of the boson-related self-consistent equations are
identically those of the $J=0$ model derived in the previous
section. The only difference arises in the numerical value of
$\chif$ which enters the overall energy scale $2\chif D$. For $J=0$
we had $\chif = (1-\delta^2)/2$, but with $J\neq 0$ it is determined
self-consistently by  Eq. (\ref{chif-eq}). Following earlier
results\cite{kotliarliu}, we anticipate that $\chif$ will remain
largely constant over a wide doping range. The critical density
$\delta_{c}$ is still given by Eq. (\ref{delta-BEC}) with
appropriate $\chif$. The incoherent-to-coherent crossover in the
electron spectral function with the vanishing of the single-particle
charge gap occurs at this density of holes. The critical doping
$\delta_c$ is plotted in Fig. \ref{phases} as empty squares. The
fermion sector has a $d$-wave symmetry gap in the spectrum
throughout the whole doping range until $\Deltaf$ itself vanishes at
a larger doping. Therefore, $\delta_c$ marks the separation of the
SLI from the nm-dSC (non-magnetic $d$-wave superconductor). The
spectral and electromagnetic properties of the individual layer is
discussed now.

The zero-temperature electronic Nambu Green's function for the
region $\delta < \delta_{c}$ is worked out as

\ba G^e (k,i\omega) &=& \sum_q  { A_{k+q} \sinh^2 \theta_q \over
E^f_{k+q} + E^b_q - i\omega} \nn
&+& \sum_q  { B_{k+q} \cosh^2 \theta_q\over E^f_{k+q} + E^b_q +
i\omega} \nn
\nn && \label{Ge-in-tJ}\ea
with $A_k$ and $B_k$ given by

\ba && A_k = \left(\begin{array}{cc} \cos^2 \theta_k & \cos\theta_k
\sin\theta_k \\ \cos\theta_k \sin\theta_k & -\sin^2 \theta
\end{array} \right) \nn
&&  B_k = \left(\begin{array}{cc} -\sin^2 \theta_k & \cos\theta_k
\sin\theta_k \\ \cos\theta_k \sin\theta_k & -\cos^2 \theta_k
\end{array} \right) \nn
&& \cos 2\theta_k = {\efk \over \Efk }, ~~ \sin 2\theta_k =
{\Deltaf_k \over \Efk } .
\ea
The corresponding spectral function contains a factor $\delta
(\omega - E^f_{k+q}-E^b_k)$ and $\delta (\omega + E^f_{k+q} + E^b_q
)$, which yields zero when $\omega = 0$, provided the boson gap
persists. Thus the spectral density is zero at $\omega = 0$ for the
SLI phase, as was the case for $J=0$. It should be noted that in
regard to the single-particle gap, we no longer have the pure
$d$-wave symmetry as was the case for a single-layer $tJ$ model. The
existence of exciton pairing renders the gap symmetry in the SLI
phase essentially that of $d+s$.

Now we present the proof that due to the exciton gap, the underdoped
region $\delta < \delta_{c}$ exhibits a superfluid response to an
external field. It is clear that the fermion sector behaves as a
superconductor with $\sigma^f = \infty$ because of the $d$-wave
pairing. It remains to show that the boson sector also has $\sigma^b
= \infty$, and we would have $\sigma^e = \infty$ according to the
Ioffe-Larkin composition rule.

The method for calculating the electromagnetic response to an
electric field for a discrete lattice model had been developed
earlier by Scalapino \textit{et al}.\cite{scalapino}. Although it
was used to calculate the response properties of a fermion system,
the formalism actually applies equally well to bosons. We obtain the
zero-temperature superfluid density

\be \rho_s (T=0) = {(V \Deltab t\chif)^2 \over 2}\sum_k {
\sin^2 k_x + \sin^2 k_y \over (\Ebk)^3 }
\label{zero-T-SF} \ee
for the upper or lower plane of the bilayer. It is non-zero as long
as the exciton pairing amplitude $\Deltab$ persists. Except for the
$(\Deltab)^2$, the rest of the terms in Eq. (\ref{zero-T-SF}) proved
to be nearly independent of doping in our numerical analysis. As a
result $\rho_s \sim (\Deltab)^2 \sim \delta$ for small $\delta$. The
exciton pairing is responsible for the superfluid behavior of the
bosons in each layer. The in-plane electron response in the
non-superconducting SLI phase is that of a superfluid because
$\sigma^b$ and $\sigma^f$ are both infinite, with $\sigma^e \sim
\delta$ at low doping.

We note that non-zero $\Deltab = \langle b_{1i}b_{2i}\rangle$
implies the total electron number in the individual layer is not
conserved; rather only the sum over the two layers is. The number
fluctuation within each layer is reminiscent of the similar
fluctuation in the superconducting ground state and is responsible
for the similar electromagnetic response.
\\

\section{t-J Model with Magnetic Order}
\label{magnetic-t-J}

The previous sections dealt with the physics of strongly interacting
bilayers in increasing degree of complexity. First we treated the
case of in-plane hopping subject only to the Mott constraint and the
inter-layer Coulomb interaction that gave rise to the exciton
formation. At the next level of complexity we introduced the
possibility of fermion pairing and of superconducting state in each
layer. In both cases we found an incoherent regime characterized by
a charge gap at low doping that we called the spin-liquid-insulator,
or SLI.

\begin{figure}[h]
\includegraphics[width=8.5cm]{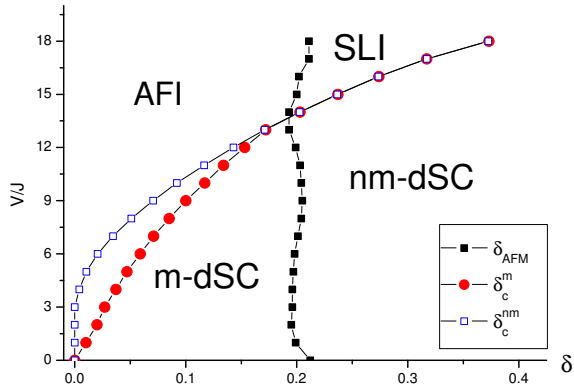}
\caption{\small (color online) $\delta_{AFM}$ (black) and
$\delta^m_{c}$ (red) for different inter-layer Coulomb interaction
$V/J$ with fixed $t/J=4$ according to the mean-field decoupling
scheme of section \ref{magnetic-t-J}. The empty blue symbols are
$\delta_{c}$ obtained with the magnetic moment suppressed to zero
(section \ref{non-magnetic-t-J}). Various phases are identified: AFI
(antiferromagnetic insulator), SLI (spin liquid insulator), m-dSC
(antiferromagnetic $d$-wave superconductor), nm-dSC (non-magnetic
$d$-wave superconductor).} \label{phases}
\end{figure}

In this section, the spectrum of possible phases in the
bilayer system is enlarged to include the antiferromagnetic
(AFM) order characterized by $m_i =(1/2)(-1)^{i}\langle
f_{ai\uparrow }^{+}f_{ai\uparrow }-f_{ai\downarrow
}^{+}f_{ai\downarrow }\rangle =m $. We consider the effects
of the magnetic order on the overall phase diagram by
decoupling  the Heisenberg exchange interaction in the
hopping, pairing, and magnetic channels. In addition to
$\delta_c$ we have the critical concentration
$\delta_{AFM}$ which marks the vanishing of magnetic order.
There are four phases found, depending on the
presence/absence of magnetism and of the charge gap. SLI
phase, which existed for arbitrary $V$ in the previous
analysis, is now only found for a sufficiently large $V/J$
and $\delta$. The low-doping region is replaced by the
antiferromagnetic insulator (AFI). The superconducting
phase is also distinguished by the co-existence of
magnetism as the m-dSC (magnetic $d$-wave superconductor)
at low doping and nm-dSC at higher $\delta$. The four
phases collide at a single tetra-critical point. Below we
present the details of the mean-field calculation.

Using the slave-boson substitution, the fermionic mean-field
Hamiltonian $H_{mf}^{f}$  becomes\cite{hanlee}

\ba H^f_{mf} &=& \sum_{ak\sigma } \efk \fbar_{ak\sigma}f_{ak\sigma}
- \sum_{ak} \Deltaf_k [f_{ak\uparrow}f_{a\kbar \downarrow} +
\fbar_{a\kbar \downarrow}\fbar_{ak\uparrow} ] \nn &&
~~~~-2Jm\sum_{k\sigma } \sigma f_{ak+Q\sigma }^{+}f_{ak\sigma}
\end{eqnarray}%
where $Q$ is the AFM wave vector $(\pi ,\pi )$. Other notations are
identical to those used in Eq. (\ref{mf-H2}). The boson mean-field
Hamiltonian remains unchanged from Sec. \ref{non-magnetic-t-J}.
Diagonalization of the fermion Hamiltonian and working out the
self-consistent equations for all the parameters involved in the
Hamiltonian can be done by straightforward
manipulation\cite{hanlee}. The self-consistent equations obtained
can be solved numerically for each doping $\delta$ and a given
inter-layer interaction strength $V$. As is well known from previous
mean-field studies\cite{hanlee}, the magnetic order vanishes at
around 18\% doping for a single layer $tJ$ model with $t/J=3$. In
our numerical solution the critical doping concentration for the
vanishing of magnetic order, $\delta_{AFM}$, does not vary greatly
with $V$.

The two critical doping concentrations, $\delta_{AFM}$ and
$\delta_{c}$, are worked out for varying strengths of $V$ in Fig.
\ref{phases} with $t/J=4$. The value of $\delta_c$ obtained in the
presence of magnetic order deviates from that obtained in the
previous section, when magnetic order was suppressed. We use
$\delta^m_c$ and $\delta^{nm}_c$ to differentiate the magnetic and
non-magnetic critical concentrations. Both are plotted in Fig.
\ref{phases}. There are two new phases which were unobserved in the
non-magnetic model, due to the co-existence of magnetism. The
charge-gapped phase, which was SLI in the non-magnetic model, is
either an AFI or SLI depending on whether the magnetic moment is
non-zero or not. The $d$-wave superconducting phase also splits into
non-magnetic (nm-dSC) and magnetic (m-dSC) $d$-wave superconducting
phases. The magnetic phase is seen to dominate the low-doping region
as expected. The spin liquid phase is pushed to high-$V/J$,
high-$\delta$ region due to a competition with antiferromagnetic
ordering, but still has a finite range of existence. The in-plane
electromagnetic response remains that of a superfluid throughout the
entire phase diagram, except in the highly doped region, not shown
in Fig. \ref{phases}, where $\Delta^b$ or $\Delta^f$ itself
vanishes.
\\

\section{Summary and Outlook}

In this paper we worked out the phase diagram and some of
the unique physical properties in a model of coupled
Mott-Hubbard bilayers. While all the phases of a single
layer model have their counterparts in the bilayer model
for a sufficiently weak inter-layer interaction strength,
there also exists a new phase with a gap in the
single-charge excitation and incoherent quasiparticle
spectra for sufficiently strong $V$ and/or $\delta$. This
spin liquid insulator phase is to be distinguished from
either the spin liquid obtained by frustrating the magnetic
order\cite{chandra}, or the metallic spin liquid state in
the high-T$_c$ phase diagram. The uniqueness of the SLI
phase we find arises from the fact that it is a charge
insulator even though the doping is finite.

On the other hand, bilayer exciton formation gives rise to
the superfluid electromagnetic response in each layer in
the charge-gapped, spin-liquid phase. Such dichotomy of the
single-particle and two-particle response functions can be
explored in future experimental setup.

It will be of importance to establish the existence of the
exotic SLI phase found in the present paper beyond the
mean-field level and applying techniques other than
slave-particle approaches. Effects of gauge fluctuations on
the mean-field phases will be considered in the future
work. Employing dynamical mean-field theory can help carve
out the phase diagram of the model Hamiltonian used in this
paper beyond the slave-particle mean-field theory.

\acknowledgments  H.J.H. thanks professor Dung-Hai Lee for
initiating the discussion of the bilayer physics during his visit to
Berkeley and Tiago Ribeiro and Alex Seidel for collaboration on a
closely related project\cite{berkeley}. This work was supported by
Korea Research Foundation through Grant No. KRF-2005-070-C00044.

\end{document}